# BULK SCHEDULING WITH DIANA SCHEDULER

*Ashiq Anjum[1, 2], Richard McClatchey[2], Arshad Ali[1], Ian Willers[3]*

1) National University of Sciences and Technology, Rawalpindi, Pakistan

2) University of the West of England, Bristol, UK

3) CERN, European Organization for Nuclear Research, Geneva, Switzerland

*Abstract*

Results from and progress on the development of a Data Intensive and Network Aware (DIANA) Scheduling engine, primarily for data intensive sciences such as physics analysis, are described. Scientific analysis tasks can involve thousands of computing, data handling, and network resources and the size of the input and output files and the amount of overall storage space allocated to a user necessarily can have significant bearing on the scheduling of data intensive applications. If the input or output files must be retrieved from a remote location, then the time required transferring the files must also be taken into consideration when scheduling compute resources for the given application. The central problem in this study is the coordinated management of computation and data at multiple locations and not simply data movement. However, this can be a very costly operation and efficient scheduling can be a challenge if compute and data resources are mapped without network cost. This can result in performance degradation particularly if no advantage is taken by a scheduling engine of recent advances in networking technologies and bandwidth abundance. We have implemented an adaptive algorithm within the DIANA Scheduler which takes into account data location and size, network performance and computation capability to make efficient global scheduling decisions. DIANA is a performance-aware as well as an economy-guided Meta Scheduler. It iteratively allocates each job to the site that is likely to produce the best performance as well as optimizing the global queue for any remaining pending jobs. Therefore it is equally suitable whether a single job is being submitted or bulk scheduling is being performed. Results suggest that considerable performance improvements are to be gained by adopting the DIANA scheduling approach.

## 1 INTRODUCTION

In scientific analysis environments such as High Energy Physics Analysis, hundreds of end-users may submit individually or collectively thousands of jobs accessing some subset of the physics data distributed over the world; this type of job submission is known as bulk submission. Currently bulk submission of jobs is employed in production (i.e. the physical process where new physics data is generated). An assignment of more than one job of a similar nature may also be made to a particular site. This is known as bulk scheduling. Given the large number of jobs that can result from this job-splitting, it should be possible to submit the job cluster to the scheduler as an unique entity, with subsequent optimization in the handling of the input sandboxes. Jobs may compete for scarce resources and this can distribute the load disproportionately among the Grid nodes. Previous approaches have been based on 'greedy' algorithms in which a job is submitted to a best resource without assessing the global cost of this action. However, this may lead to a skewed distribution of resources resulting in large queues and performance and throughput degradation for the remainder of the jobs. We present a scheduling system which not only allocates best available resources to a job but also checks the global state of jobs and resources so that the strategic output of the Grid is maximized and no single job can undergo starvation.

In this paper we describe the so-called DIANA Scheduling system and in particular its usage in scheduling bulk jobs. We discuss network issues in scheduling data intensive jobs and present an algorithm for scheduling bulk jobs. We illustrate that a priority driven multi-queue feedback based approach is the most feasible to tackle the issue of bulk scheduling.

## 2 DIANA SCHEDULING

Data intensive applications often analyze large amounts of data which are replicated over geographically distributed sites. If data are not replicated to the site where the job is intended to be executed, the data need to be fetched from remote sites. This data transfer from other sites can degrade the overall performance of the job execution. If a computing job runs at a remote site, the output data produced needs to be transferred to the user so he can analyse the result locally. For performance gains in the overall job execution time and to maximize the Grid throughput, we need to align and co-schedule the computation and the data (input as well as output) in such a way that we can reduce the overall computation and data transfer costs. We may even decide to send both the data and the executables to a third location depending on the capabilities and characteristics of the computing, network and storage resources.

We not only need to use the network characteristics while aligning data and computations, but we also need to optimize the task queues of the Meta-Scheduler on the basis of this correlation so that network characteristics can play an important role in the matchmaking process and on Grid scheduling optimization. Thus, a more complex scheduling algorithm is required that should consider the job execution, the data transfer and their relation with various network parameters on multiple sites. Hence there are three core elements of the scheduling problem which need to be tackled and we express each of them as a separate cost in the DIANA scheduling algorithm. Data Location, network capacity/quality and the available computation cycles are the three major elements which need to be incorporated in the scheduling decisions.

First we calculate the network cost. TCP is a closed loop protocol in which packet delivery is acknowledged; it sends a number of packets (called a Window) in a Round Trip Time (RTT). The larger the numbers of packets that can be sent in a Window, the larger the connection rate. Thus:

Throughput = Average_Window/Average_RTT

Where Average Window is inversely proportional to the square-root of loss probability, i.e. the greater the losses the lower the Average Window. Losses are dependent on path conditions [1] and therefore Network cost is:

Network Cost=Losses/Bandwidth

The second important cost which needs to be part of the scheduling algorithm is the computation cost. [2] describes a mathematical formula to compute the processing time or Compute Cost of a job:

$$\frac{Q_i}{P_i} \times W_5 + \frac{Q}{P_i} \times W_6 + SiteLoad \times W_7$$

Where $Q_i$ is the length of the waiting queue, $P_i$ is the computing capability of the site i and SiteLoad is the current load on that site. $W_5$, $W_6$ and $W_7$ are weights which can be assigned depending upon the importance of the queue and the processing capability. The third most important cost aspect in data intensive scheduling is the data transfer cost::

Data Transfer Cost (DTC) = Input Data Transfer Cost + Output Data transfer cost + Executables transfer cost

Here we take three different costs for data transfer. The input data transfer cost is the most significant since most jobs take large amounts of input data which depends further on the network cost. Higher network cost will increase the data transfer cost and vice versa. The same is the case for output data since output data needs to be transferred to the user. Once we have calculated the cost of each stake holder, the total cost is simply a combination of these individual costs thus:

Total Cost = Network Cost + Computation Cost + DTC

The main optimization problem that we want to solve is to calculate the cost of data transfers betweens *sites* (DTC), to minimize the network traffic cost *between the sites* (NTC) and to minimize the computation cost of a job *within a site*. This total cost covers all aspects of the job scheduling and gives a single value for each associated cost, thus optimizing the Meta-scheduling decisions.

## 3 PRIORITY AND BULK SCHEDULING

We describe here a few Characteristics which can help us in creating an optimized scheduling algorithm as described later in section 4. CPU Utilization is the first performance criterion. We want to keep the CPU as busy as possible. If the CPU is busy in exchanging processes, then work is being carried out. One measure of work is the number of jobs completed per unit time called the throughout. The interval from the time of submission to completion is termed the turnaround time and has significant bearing on performance indicators. Turnaround time is the sum of the periods spent waiting to access memory, waiting in the ready queue, executing the CPU and doing input/output. The waiting time is the sum of the periods spent waiting in the ready queue. In an interactive system, turnaround time may not be the best criterion. Another measure is the time from the submission of a request until the first response has been provided. This measure, called response time, is the time it takes to start responding but not the time that it takes to output that response. In the proposed algorithm, we tend to maximize CPU utilization and throughput and minimize turnaround time, waiting and response time.

### 3.1 PRIORITY BASED SCHEDULING

The proposed scheduling algorithm described later is termed a priority algorithm. A priority is associated with each process and the CPU is allocated to the process with the highest priority. Equal priority processes are scheduled on a First Come First served (FCFS) basis. We discuss scheduling in terms of high priority and low priority. Priorities can be defined either internally or externally. Internally defined priorities use some measurable quantity or quantities to compute the priority of a process.

For example, time limits, memory requirements, the number of open files and the ratio of I/O to CPU time can be used in computing priorities. External priorities are set by criteria that are external to the scheduling system such as the importance of the process. Priority scheduling can be either pre-emptive or non pre-emptive. The bulk scheduling algorithm described here is not a pre-emptive one. It simply places the new job at the head of the ready queue and does not abort the running job.

### 3.2 MULTILEVEL QUEUE SCHEDULING

Due to the different quality of service requirements by the community of Scientific Analysis users, jobs can be classified into different groups. For example, a common division is made between interactive jobs and batch jobs. These two types of jobs have different response-time requirements, and so might have different scheduling needs. In addition, interactive jobs may have priority over batch jobs. A multilevel queue-scheduling algorithm partitions the ready queue into multiple separate queues as shown in Figure 1.

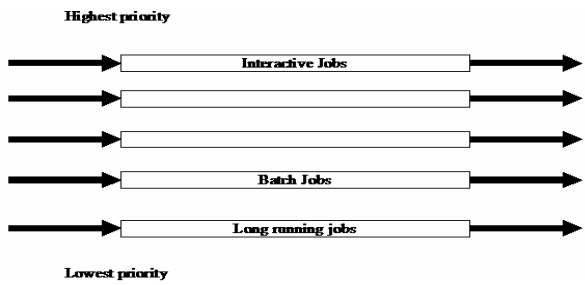

**Fig1** Multi level queue scheduling

The jobs are assigned to one queue, generally based on some property of the process, such as memory size, process priority, or process type. Each queue has absolute priority over lower-priority queues. No job in the batch queue for example, can run unless the queues for system processes, interactive processes, and interactive editing processes are all empty.

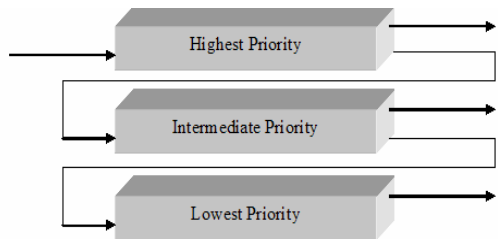

**Fig2** Multilevel feedback queues

In a multilevel queue-scheduling algorithm, jobs are permanently assigned to a queue on entry to the system. Jobs do not move between queues and this can create starvation if the jobs running are long duration jobs. We have employed multilevel feedback queue scheduling as shown in Figure 2 since it allows a job to move between queues. The idea is to separate processes with different requirements and priorities. If a job uses too much CPU time or is very data intensive, it will be moved to a higher-priority queue. Similarly, a job that waits too long in a lower-priority queue may be moved to a higher-priority queue.

## 4 BULK SCHEDULING ALGORITHM

By following all of the above points we propose a multilevel feedback queue and priority-driven scheduling algorithm for bulk scheduling. The salient features of this bulk scheduling algorithm are now briefly discussed.

1. Job priority is important while scheduling the bulk jobs. High priority jobs are executed first. The priority of jobs starts decreasing if the number of jobs from a user/site increases beyond a certain point. It becomes less than all the jobs in the queue if job frequency is very high.
2. A priority scheduling algorithm may leave some low priority processes waiting indefinitely for the CPU (so-called starvation). We use an aging technique to overcome this problem. Aging is a technique of gradually increasing the priority of processes that have been waiting in the system for a significant time. Starvation of the resources is controlled by controlling the priority of the jobs. If no other job is available in the queue then all jobs from the user/site will be executed as high priority jobs. We do not employ quota and accounting since this restricts the users to a particular limit. Instead we use priority to schedule bulk jobs and to control the frequency as well as the queue on this basis. Similarly we do not follow the budget and deadline method of economy-based scheduling since the Grid is dynamic and volatile and deadline is feasible only for static type of environment.
3. All of the bulk jobs in a single burst will be submitted at a single site. If data and computing capacity is available at more than one site, we can think of job splitting and partitioning. Queue length, Data Location, load and network characteristics are key parameters for making scheduling decisions for a site. The priority of the burst or bulk of jobs is always the same since each batch of jobs has the same execution requirements.
4. We do not use a pre-emptive scheduling approach. Rather due to the interactive nature of most of the jobs, we follow a non pre-emptive mode of scheduling and execution. Since most jobs are data intensive, this makes it increasingly important to consider the non pre-emptive mode as a primary approach.
5. A 'Round Robin' approach inside queues is not feasible in this case since most of the analysis jobs are interactive and the user is eagerly awaiting the output. Any delay in the output may lead to a disatisfied user and this requires us to provide resources until the output can be seen. This also leads to the conclusion that the pre-emptive approach is not feasible for interactive jobs but can be considered for batch jobs. In this algorithm we consider only the interactive jobs used for a Grid-enabled analysis. Job migration between priority queues is a key point of the algorithm. Jobs can move between low priority to high priority queues depending upon the number of jobs from each user and the time passed in a particular low priority queue. Although migration of the jobs between queues is supported within a single queue, we use FCFS algorithm. Before jobs are placed inside the queue for execution, the algorithm arranges the jobs using the Shortest Job First (SJF) algorithm. We use the number of processors required as a criterion to decide between short or long time. Fewer processors required means job execution time is shorter and its priority should be set higher. All shorter jobs are executed before longer jobs; this reducing the average execution time of all jobs.
6. Priorities can be of three types: user, quota and system centric. We employ a system centric policy (embedded inside the scheduler) since otherwise users can manipulate the scheduling process. In this way a uniform approach will be set by the scheduler for all users and a similar priority will be applied to all stake holders.
7. Knowing the job arrival rates and execution capacity, we can compute utilization, average queue length, average wait time and so on. As an example, let N be the average queue length (excluding the jobs being serviced), let W be the average waiting time in the queue, and let R be the average arrival rate for new jobs in the queue. Then, we expect that during the time W that a job waits, R*W new jobs will arrive in the queue. If the system is in a steady state, then the number of jobs leaving the queue must be equal to the number of jobs that arrive.

$$N = R*W$$

This equation, known as Little's Formula [3], is valid for any scheduling algorithm and arrival distribution. We can use it to compute any one of the three variables, if we know the other two. If the arrival rate of the jobs is more than the capacity of the site to compute, then we export jobs to some other site which is least loaded and can compute the results within less time than the current site.

8. When a site is assigned too many jobs, it can try to send a number of them to other sites, which have more free resources or are processing fewer jobs than the local site at that time; in this case, the jobs move from one site to another based on the criteria described in section 2. The scheduler queries all the sites for their average load at that time. The one with the minimum cost will be selected. Once a job has been submitted on a remote site, the site at which it arrives will not attempt to schedule it again on some other remote site (thus avoiding the situation in which a job cycles from one site to another). To each site we submit a number of jobs and a job reads an amount of data from a local database server, and then processes the data. If a site becomes loaded and jobs need to be scheduled on a remote site, the cost of their execution increases as the database server is no longer at the same site. If the amount of data to be transferred is too large or the speed of the network connections is too low, it might be better not to schedule jobs to remote sites but to schedule them for local execution.

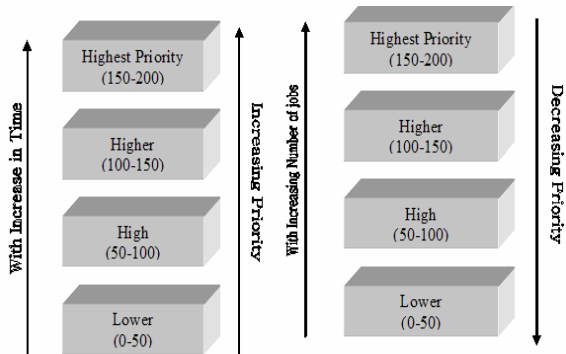

**Fig3**     Priority with Time and Job Frequency

9. In bulk scheduling there is a time threshold and a job threshold. If the number of jobs submitted from a particular user increases beyond the job threshold then the priority of the jobs submitted above the threshold number is decreased and jobs are migrated to a lower priority queue. In other words, with an increasing number of jobs, the priority of jobs from a particular user starts to decrease. Moreover, a time threshold is included to reduce the aging affect. With the passage of time, the priority of jobs in the lower priority queues is increased so that it can also have a chance of being executed after a certain wait time. In other words, the more time a job has to wait the more its priority continues to increase. This is illustrated in figure 3.

## 5     RESULTS AND DISCUSSION

We have presented a set of results which we have observed through implementation of the DIANA scheduler using MONARC [4] simulations to check the algorithm behaviour for bulk scheduling. First we submitted a number of jobs greater than the processing capacity of the site and observed large queues of jobs which cannot be processed in an optimal manner. The bulk scheduling algorithm discussed above was used to move the jobs to other sites. Shown below is the summary of the jobs which were executed locally and those migrated to other sites. The results suggest that as the number of jobs increases beyond the threshold limit, more and more jobs are migrated to other less loaded sites over time since the current site selection is no longer optimal.

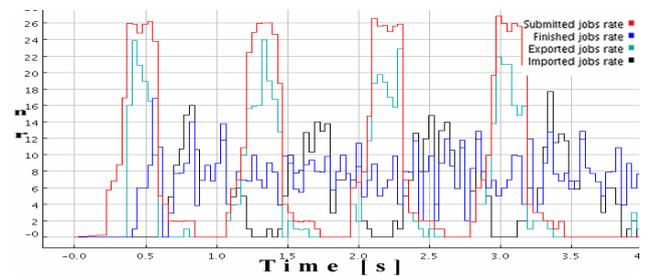

**Fig4**   Jobs execution and migration with Time

Once jobs at a site cross the threshold limit, the DIANA scheduling algorithm is used to select the best alternative site for execution in terms of computation power, data location, network capacity and queue length. The following graph shows the optimization achieved by employing the DIANA algorithm.

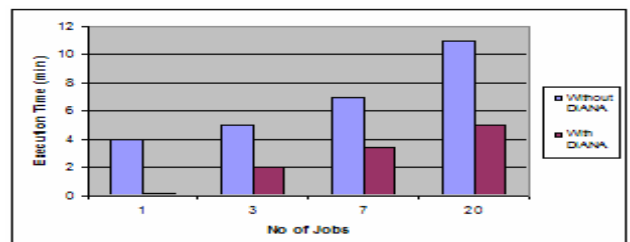

**Fig5**     Execution time versus number of jobs

Results indicate that considerable optimization can be achieved using bulk scheduling and DIANA scheduling algorithms.